\documentclass[twocolumn,a4paper,nofootinbib,
showpacs,aps,floatfix]{revtex4}
\usepackage{graphicx}
\usepackage{bm}
\usepackage{amsmath}
\def\la{\langle}
\def\ra{\rangle}

\begin {document}
\title{Vibrational Bloch-Siegert effect in trapped ions}
\author{I. Lizuain}
\email[Email address: ]{ion.lizuain@ehu.es}
\affiliation{Departamento de Qu\'\i mica-F\'\i sica,
Universidad del Pa\'\i s Vasco, Apdo. 644, Bilbao, Spain}
\author{J. G. Muga}
\email[Email address: ]{jg.muga@ehu.es} 
\affiliation{Departamento de Qu\'\i mica-F\'\i sica,
Universidad del Pa\'\i s Vasco, Apdo. 644, Bilbao, Spain} 
\author{J. Eschner}
\email[Email address: ]{juergen.eschner@icfo.es}
\affiliation{ICFO, Institut de Cienci\`es Fot\`oniques, 08860 Castelldefels, Barcelona, Spain}


\begin{abstract}
When trapped atoms are illuminated by weak lasers, off-resonant 
transitions cause shifts in the frequencies of the vibrational-sideband 
resonances. These frequency shifts may be understood in terms of 
Stark-shifts of the individual levels or, as proposed here, as a 
vibrational Bloch-Siegert shift, an effect closely related to the usual 
(radio-frequency or optical) Bloch-Siegert shift and associated with rapidly 
oscillating terms when the Rotating Wave Approximation is not made. 
Explicit analytic expressions are derived and compared to numerical 
results, and the similarities and differences between the usual and the 
vibrational Bloch-Siegert shifts are also spelled out.
\end{abstract}
\maketitle
\section{Introduction}
\subsection{Bloch-Siegert shift}
The Rotating Wave Approximation (RWA) is applied to describe atoms interacting with near-resonant fields. It consists on neglecting the rapidly oscillating counter-rotating terms in a Hamiltonian.
In 1940 Bloch and Siegert, studying magnetic resonances, showed that if the RWA is not applied, 
the counter rotating, fast oscillating terms give rise to a shift in the resonance frequency of the magnetic dipoles, 
i.e., the Bloch-Siegert shift \cite{BS40,AllenEberly,WWM97}.  
Similar shifts exist in principle in the optical domain, although much smaller and difficult to detect. 

In the treatment of laser driven trapped ions, apart from an \emph{optical} RWA for the driven optical transition, 
a second or ``vibrational 
RWA'' is usually applied \cite{LBMW03}. Within this approximation, 
the absorption spectrum of a harmonically trapped (two-level) ion consists of a carrier band centered at the transition
frequency $\omega_0$ and sidebands separated from the carrier by multiples of the trap frequency $\omega_T$.
If this second RWA is not applied and vibrational counter-rotating terms are taken into account, the energy levels
are distorted and thus the position of the sideband resonances are shifted, an effect that we shall call a ``vibrational 
Bloch-Siegert shift''.
Its analysis is the objective of the present paper in which we shall 
provide explicit expressions within a perturbative approach based on the 
resolvent method. The peculiarities 
of the Hamiltonian for the trapped ion lead to differences between the 
ordinary (in magnetic or optical resonances) and vibrational BS effects that we shall also discuss.        

These sideband frequency shifts are important for optimizing the operation of 
quantum gates based on laser-driven trapped ions and can be observed and compensated experimentally 
\cite{steane00,SchmidtKaler03}.
In this context they have been understood as the result of Stark-shifts of the individual levels due to off-resonant transitions.    
The connection between our compact treatment of the frequency shift (as a vibrational Bloch-Siegert 
effect) and the Stark-shifts will also be spelled out. Thanks to the systematic treatment with the resolvent method 
we find general expressions for the frequency shifts of arbitrary sidebands and
correction terms that had been previously overlooked.

\subsection{Vibrational Rotating Wave Approximation}
\label{vibrationalRWA}

We consider a two-level ion with (internal) ground and excited states $|g\ra$ and $|e\ra$, and transition frequency $\omega_0$ among them, moving in an effective one-dimensional harmonic potential of frequency $\omega_T$ in $x$-direction (motion in $y$ and $z$ directions is unexcited
and ignored throughout); this ion is illuminated by a (classical) laser beam with traveling wave of wavevector in $x$-direction and wavenumber $k_L$.    
The system is described in the dipole approximation 
by
\begin{equation}
\label{starting_hamiltonian}
H(t)=H_T+H_A+\hbar\Omega_R\left(\sigma_++\sigma_-\right)\cos\left(\omega_Lt-k_Lx\right),
\end{equation}
with
\begin{eqnarray}
\label{H_trap}
H_T&=&\hbar\omega_T a^\dag a,\\
\label{H_atom}
H_A&=&\frac{\hbar\omega_0}{2}\sigma_z.
\end{eqnarray}
where $\sigma_z=|e\ra\la e|-|g\ra\la g|$, $\sigma_+=|e\ra\la g|$, and $\sigma_-=|g\ra\la e|$; 
$\Omega_R$ is the on-resonance Rabi frequency, which plays the role of an atom-field coupling constant, and 
$a^\dag$ ($a$) are the creation (annihilation) operators of the harmonic potential.

In an interaction picture defined by $H_T+H_A$ and neglecting the fast oscillating terms within the usual (optical) RWA,
the Hamiltonian becomes 
\begin{eqnarray}
\label{time_dep_ham_01}
H_{TA}(t)&=&\frac{\hbar\Omega_R}{2}\left( e^{i\eta\left[a(t)+a^\dag(t)\right]}
e^{-i\Delta t}\sigma_+
 + H.c.\right),
\end{eqnarray}
where $H.c.$ means ``Hermitean conjugate'', $a(t)=a e^{-i\omega_Tt}$, $a^\dag(t)=a^\dag e^{i\omega_Tt}$, and $\Delta=\omega_L-\omega_0$ is the detuning between the laser frequency and the internal atomic transition frequency. 
The parameter $\eta=k_Lx_0$ is known as the Lamb-Dicke (LD) parameter,
where $x_0=\sqrt{\hbar/2m\omega_T}$ is the extension (square root of the variance) of the ion's ground state, i. e., 
$x=x_0(a+a^\dag)$.
The LD parameter is a measure of the 
trap width on the scale of the laser wavelength.
Let us denote by $|g,n\ra$ ($|e,n\ra$) the state of the ion in the ground (excited) internal state and in the $n$-th 
motional level of the harmonic oscillator. 
Using the so called BCH identity \cite{Orszag},  
\begin{equation}
\label{BCH}
e^{i \eta\left[a(t)+a^\dag(t)\right]}\equiv e^{-\frac{\eta^2}{2}}e^{i\eta a^\dag(t)}e^{i\eta a(t)}, 
\end{equation}
and expanding the exponentials in power series of $\eta$, we end up with a Hamiltonian with
terms containing a combination of $\sigma_{\pm}$, with $n$ $a^\dag$-operators and $n'$ $a$-operators rotating with
a frequency $(n-n')\omega_T$ \cite{LBMW03},
\begin{eqnarray}
\label{time_dep_ham_02}
H_{TA}(t)\!\!&\!\!=\!\!&\!\!\frac{\hbar\Omega_R}{2}
\!\!\left[\!e^{-\frac{\eta^2}{2}}\sum_{nn'}
\frac{\left(i \eta\right)^{n+n'}}{n!n'!}\!  a^{\dag n}\! a^{n'}
e^{i (n-n') \omega_T t}\!e^{-i\Delta t}\sigma_+\right.
\nonumber\\
&+& {H.c.}\Bigg] 
\end{eqnarray}
which in principle couples all the different vibrational levels via de $a^{\dag n}$ and $a^{n'}$ operators.
The combinations satisfying the  $\Delta\approx k\omega_T$ ($k=n-n'$) condition will be resonant, coupling the
states $|g,n\ra\leftrightarrow|e,n+k\ra$, while the rest of rapidly
oscillating terms are usually neglected in a second application of the RWA, a vibrational RWA (VRWA) \cite{LBMW03}.
Within this approximation,  only
co-rotating states $|g,n\ra\leftrightarrow|e,n+k\ra$ are coupled  and the system becomes two-dimensional at 
each of these resonances. They are classified as:
%
(i) $\Delta\sim0$: Carrier resonance. The system is described 
by a Hamiltonian which is equivalent to a two level atom in a resonant field,
coupling the states
$|g,n\ra\leftrightarrow|e,n\ra$; 
%
(ii) $\Delta\sim-k\omega_T$: $kth$ red sideband. The system is described
by a $k$-photon Jaynes-Cummings type Hamiltonian, 
which couples $|g,n\ra\leftrightarrow|e,n-k\ra$; 
%
(iii) $\Delta\sim k\omega_T$: $kth$ blue sideband.
The system is described
by a $k$-photon anti Jaynes-Cummings type Hamiltonian, 
coupling the states $|g,n\ra\leftrightarrow|e,n+k\ra$.
%

The differences and similarities of both approximations (RWA and VRWA) are discussed in detail in Section \ref{RWAvsVRWA}.

%
\section{Vibrational Bloch-Siegert shift}
\begin{figure}[t]
\begin{center}
\vspace{1cm}
\includegraphics[height=6cm]{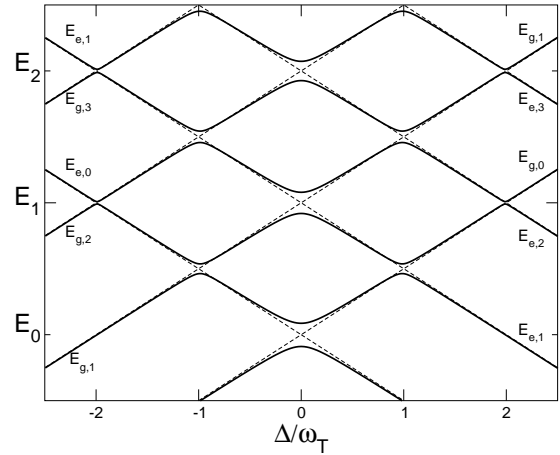}
\caption[]{Bare ($\Omega_R=0$, dashed line) and dressed ($\Omega_R/\omega_T=0.3$, solid line) energy levels
(in arbitrary units) as a function of the laser detuning, with $E_n=n\hbar\omega_T$ being the energy levels
of the trapping potential. A not too small LD parameter $\eta=0.4$ has been 
intentionally chosen in order to highlight the higher order avoided crossings. }
\label{energy_levels}
\end{center}
\end{figure}

If the VRWA is not applied and vibrational counter-rotating terms in the Hamiltonian (\ref{time_dep_ham_02}) are 
taken into account, a shift in the apparent position of each sideband resonance is observed, 
an effect that may be compared to the standard Bloch-Siegert shift in magnetic or optical resonances.  
This shift is  more easily visualized 
and calculated if one writes the original Hamiltonian
(\ref{starting_hamiltonian}) in a 
frame rotating with the field frequency, i.e., in a field-adapted interaction picture defined by the zeroth order Hamiltonian 
$\frac{\hbar}{2}\omega_L\sigma_z$. In this frame, and after applying the usual (optical) RWA, the Hamiltonian becomes time-independent,
\begin{equation}
\label{time_ind_ham}
H_{LA}=\hbar\omega_Ta^\dag a-\frac{\hbar\Delta}{2}\sigma_z+\frac{\hbar\Omega_R}{2}\left[
e^{i\eta(a+a^\dag)}\sigma_++{H.c.}\right].
\end{equation}
The energy levels of this Hamiltonian are plotted in Fig. \ref{energy_levels} as a function of the detuning.
%
%
The bare ($\Omega_R=0$) levels are given by straight lines
\begin{eqnarray}
\label{bare_energies}
E_{g,n}^{(B)}&=&n\hbar\omega_T+\frac{\hbar\Delta}{2},\\
E_{e,n}^{(B)}&=&n\hbar\omega_T-\frac{\hbar\Delta}{2},
\end{eqnarray}
which are degenerate at each of the resonances mentioned 
before (dashed lines in Fig. \ref{energy_levels}), but the
degeneracies are removed and become avoided crossings when the laser is turned on 
(solid lines in Fig. \ref{energy_levels}).
These anti-crossings are nothing but the different sidebands mentioned
in the previous section and lead to transitions between the different 
bare states. 
In this time independent approach, the VRWA corresponds to restricting  
the theory to a subspace
spanned only by the two states involved in a given anti-crossing neglecting
the rest of states, provided that the 
energy splitting of the avoided-crossing is well isolated \cite{LM07_b}. 

\begin{figure}[t]
\begin{center}
\vspace{1cm}
\includegraphics[height=6cm]{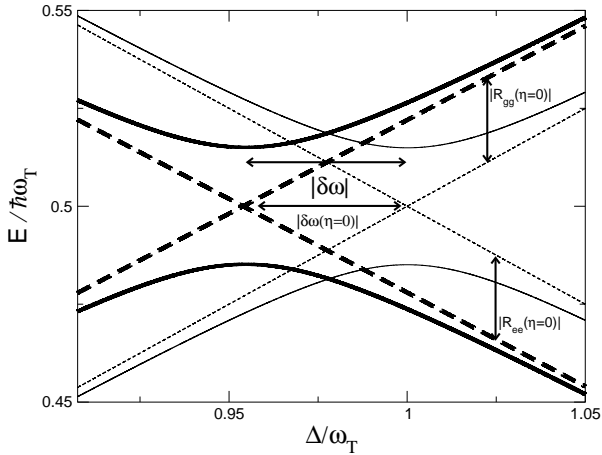}
\caption[]{Detail of the energy level diagram at the first blue sideband  between states $|g,0\ra$ and $|e,1\ra$. 
The bare ($\Omega_R=0$) energy levels cross each other since the involved states
are not coupled by the laser (thin-dashed lines). 
When the laser is turned on ($\Omega_R/\omega_T=0.3$) and the VRWA applied, 
the dressed energy levels form an avoided crossing, but the position of the resonance remains unchanged at 
$\Delta_0=\omega_T$ (thin-solid line). If the VRWA is not made and counter rotating terms are kept, the position of the
resonance is shifted to $\Delta=\Delta_0+\delta\omega$ (thick solid line). 
In these cases a LD parameter of $\eta=0.1$ has been 
used. If the laser is kept turned on but $\eta=0$ (the energy levels of the harmonic potential do not 
couple), the states $|g,0\ra$ and $|e,1\ra$ are decoupled to all orders in the perturbation. 
In this case, the energy levels cross each other while still 
being shifted by the off-resonant carrier transition (thick-dashed line).}
\label{zoomed_level_fig}
\end{center}
\end{figure}

Nevertheless, a careful treatment of the $kth$ sideband
(anti-crossings) in which all bare states are included, reveals 
that the frequency of the resonance does
not exactly coincide with the predicted value $\Delta=k\omega_T$, but 
is shifted by $\delta\omega$ due to the presence of non-resonant vibrational terms, see Fig. \ref{zoomed_level_fig}. 
In order to account for this effect one has to 
use a theory which includes not only the two main energy levels involved in each resonance, but also non-resonant terms. We shall use the resolvent method here as described in \cite{CDG98,cohen73}. 
%
%
\subsection{Perturbative treatment}

In the low intensity field approximation ($\Omega_R\ll\omega_T$), the Hamiltonian (\ref{time_ind_ham}) 
describing the system may be written as a sum of a bare (unperturbed) part $H_B$ 
and a (small) perturbative term $V$,
\begin{eqnarray}
\label{bare_ham}
H_{B}&=&\hbar\omega_T a^\dag a-\frac{\hbar\Delta}{2}\sigma_z,
\\
V(\Omega_R)&=&\frac{\hbar\Omega_R}{2}\left[
e^{i\eta(a+a^\dag)}\sigma_++H.c.\right].
\label{perturbation}
\end{eqnarray}
The bare states (eigenstates of $H_B$) of the system satisfy 
\begin{equation}
H_B|\alpha,n\ra=E_{\alpha,n}^{(B)}|\alpha,n\ra,
\end{equation}
where the index $\alpha=g,e$ accounts for the internal atomic state and the bare energy levels 
$E_{\alpha,n}^{(B)}$ are given in Eq. (\ref{bare_energies}).

Let us now consider the $|g,n_g\ra\leftrightarrow|e,n_e\ra$ sideband transition, which corresponds to the crossing between 
the bare levels $E_{g,n_g}^{(B)}$ and $E_{e,n_e}^{(B)}$ at the point
\begin{eqnarray}
\label{crossing_point}
E_0&=&\frac{\hbar\omega_T}{2}(n_g+n_e),\\
\Delta_0&=&(n_e-n_g)\omega_T
\end{eqnarray}
of the $(E,\Delta)$ plane, see Fig. \ref{energy_levels}. 
If these levels are close to each other but far from other levels, the time evolution of the system in 
the subspace spanned by the states $|g,n_g\ra$ and $|e,n_e\ra$ may be approximately described by an approximate Hamiltonian
\cite{CDG98}
\begin{equation}
\label{Heff}
\tilde H=\left(  \begin{array}{cc}
E_{g,n_g}^{(B)}+R_{gg}(E_0) & R_{ge}(E_0)\\
R_{eg}(E_0) & E_{e,n_e}^{(B)}+R_{ee}(E_0)
              \end{array}\right),
\end{equation}
where $R_{\alpha\beta}(E_0)=\la\alpha,n_\alpha|R(E_0)|\beta,n_\beta\ra$ 
are the matrix elements of the level shift operator defined by
\begin{equation}
\label{level_shift_operator}
R(E_0)=PVP+\sum_{n=1}^{\infty}PV\left(\frac{Q}{E_0-H_B}V\right)^nP,
\end{equation}
with $P=|g,n_g\ra\la g,n_g|+|e,n_e\ra\la e,n_e|$ and $Q=1-P$. 
Even though 
it is 2-dimensional, the approximate Hamiltonian (\ref{Heff}) contains information about all the non-resonant states via the operator $Q$, the projector onto the
counter-rotating subspace. 
The eigenvalues of $\tilde H$ will then give, for weak fields, the correct energy levels, including non-resonant effects.\footnote{The resolvent method is capable of providing the exact levels substituting $E_0$ by the unknown eigenvalue $E$ in  the approximate Hamiltonian. Finding $E$ requires iterative procedures, see e.g. \cite{cohen73}.
The simplest, explicit, rather than implicit, approximate treatment followed here provides the correction to the energy in leading order in the perturbation.} 
It can be easily shown\footnote{Solve $dE_{\alpha,n_\alpha}/d\Delta=0$, 
where $E_{\alpha,n_\alpha}$ are the eigenvalues of $\tilde H$.} that the maximum (or minimum) of these perturbed energy levels
are shifted from their non-perturbed position $\Delta_0$ by
\begin{equation}
\label{BS_shift}
\delta\omega=[R_{ee}(E_0)-R_{gg}(E_0)]/\hbar,
\end{equation}
which is the vibrational Bloch-Siegert shift of the position of the resonances
due to the effect of counter rotating (vibrational) terms, see Fig. \ref{zoomed_level_fig}.

Moreover, if the states are coupled by $R_{ge}$, their corresponding levels form an avoided crossing with 
an energy splitting given by 
\begin{equation}
\label{splitting}
\delta\epsilon=\left|R_{ge}(E_0)\right|=\frac{\hbar\left|\Omega_{n_g,n_e}\right|}{2},
\end{equation}
where $\Omega_{n,n'}=\Omega_R\la n| e^{i\eta(a+a^\dag)}|n'\ra$ are the coupling strengths between the different
vibrational levels of the trap \cite{WI79,wineland98}, 
see the explicit expressions in Appendix \ref{matrix_elements_calculation}. 
As we have already pointed out, the approximate Hamiltonian (\ref{Heff})
will only be valid if the anti-crossing is well isolated. A criterion for resonance isolation 
is $\delta\epsilon\ll\omega_T$, since $\omega_T$ is the energy difference between  consecutive resonances.
As low intensity  lasers ($\Omega_R\ll\omega_T$) are being assumed, this criterion will be readily 
fulfilled and all the crossings will be well isolated.

\begin{figure}[t]
\begin{center}
\vspace{1cm}
\includegraphics[height=6cm]{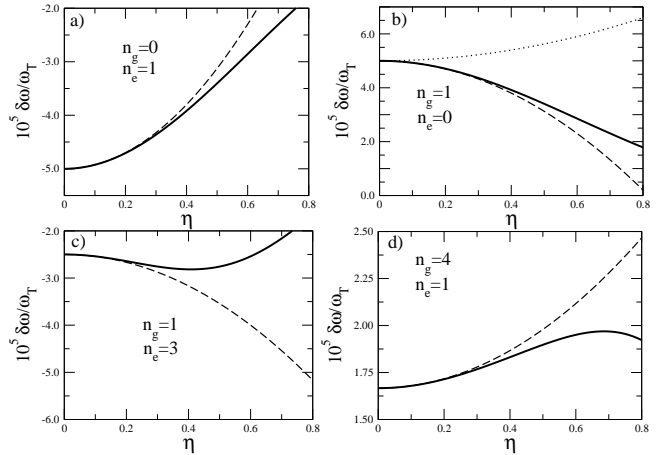}
\caption[]{Vibrational Bloch-Siegert Shift in units of the trap frequency as a function of the LD parameter $\eta$ for
different $|g,n_g\ra\leftrightarrow|e,n_e\ra$ transitions and a ratio $\Omega_R/\omega_T=0.01$.
The solid line is a virtually exact result obtained
by diagonalizing the full 
Hamiltonian (with a large basis of bare states) 
and numerically finding the maximum or minimum of the corresponding energy level.
This line is 
indistinguishable from the one obtained by computing expression (\ref{VBS}) to all orders in $\eta$.
The dashed lines correspond to the simplified expression in the LD regime, Eq. (\ref{VBS_LD}). 
The dotted line in (b) corresponds to the shift derived in \cite{steane00} and \cite{aniello04}.}
\label{vbs_eta_fig}
\end{center}
\end{figure}

In order to give an explicit expression of $\delta\omega$ we need the matrix elements $R_{gg}$ and $R_{ee}$ of the 
level shift operator.
To lowest order in the perturbation, they are given by (see Appendix \ref{matrix_elements_calculation} for the explicit calculation)
\begin{eqnarray}
R_{gg}(E_0)&=&\la g,n_g|R|g,n_g\ra
=\sum_{\substack{k=0\\k\neq n_e}}^\infty\frac{\left|\hbar\Omega_{n_g,k}/2\right|^2}{E_0-E_{e,k}^{(B)}},\nonumber\\
R_{ee}(E_0)&=&\la e,n_e|R|e,n_e\ra
=\sum_{\substack{k=0\\k\neq n_g}}^\infty\frac{\left|\hbar\Omega_{n_e,k}/2\right|^2}{E_0-E_{g,k}^{(B)}}.
\label{lso_matrix_elements}
\end{eqnarray}
Then, using the expressions of the bare energies (\ref{bare_energies}) and the crossing point of these bare levels
(\ref{crossing_point}) one has from (\ref{BS_shift}) that the position of the $|g,n_g\ra\leftrightarrow|e,n_e\ra$
resonance is given by $\Delta=\Delta_0+\delta\omega$,
where 
\begin{eqnarray}
\label{VBS}
\delta\omega
&=&\frac{1}{4\omega_T}\left(\sum_{\substack{k=0\\k\neq n_g}}^\infty\frac{\left|\Omega_{n_e,k}\right|^2}{n_g-k}-
\sum_{\substack{k=0\\k\neq n_e}}^\infty\frac{\left|\Omega_{n_g,k}\right|^2}{n_e-k}\right),
\end{eqnarray}
which is in principle valid for all values of $\eta$ provided low intensity fields are assumed.
We have verified the validity of this expression by comparing  
with the frequency shifts that result from 
diagonalizing the time independent Hamiltonian (\ref{time_ind_ham}) with a large basis of bare states ($n\gg1$) and numerically 
finding the maximum or minimum of a given energy level, see Fig. \ref{vbs_eta_fig} (solid lines).

Note that Eq. (\ref{VBS}) can also be understood as the sum of the Stark shifts for the excited state minus the Stark shifts for the ground state due to all non-resonant transitions, which provides a connection with  
previous treatments \cite{steane00,SchmidtKaler03}. 

%
%
\begin{figure}[t]
\begin{center}
\vspace{1cm}
\includegraphics[height=7cm]{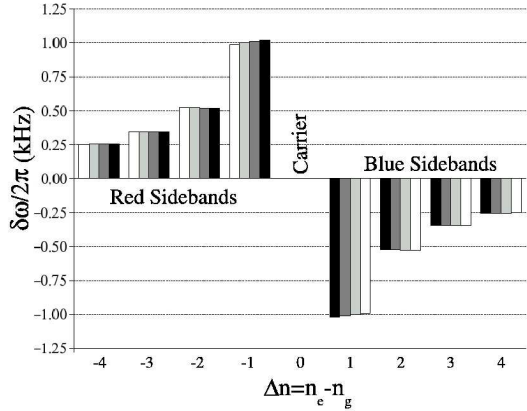}
\caption[]{Bloch-Siegert shift for the first few sidebands, with experimental data taken from \cite{Riebe06} (see text). 
The grey scale indicates the different vibrational levels
involved in a particular sideband. Black, dark grey, light grey and white correspond, respectively, 
to $n=0,1,2,3$. $n=n_g$ for blue sidebands and $n=n_e$ for red sidebands. Note the symmetry
between blue and red sidebands.    
The shift is zero for the carrier.}
\label{barplot}
\end{center}
\end{figure}

%

\subsection{LD regime: $\eta\ll1$}
A particularly interesting and common regime is the so-called
LD regime, in which the recoil frequency of the ion is much smaller than the trapping frequency, i.e, $\eta\ll1$. 
Up to quadratic terms in $\eta$, only the same or consecutive vibrational levels are coupled,
reducing the infinite sum in (\ref{VBS}) to one where only ``off-resonant carrier terms''
($k=n_e$ in $R_{ee}$ and $k=n_g$ in $R_{gg}$)
or coupling between adjacent vibrational levels is considered. 
Thus, we find that, see Fig. \ref{vbs_eta_fig} (dashed lines) 
\begin{eqnarray}
\label{VBS_LD}
\delta\omega&=&\frac{\Omega_R^2}{2(n_g-n_e)\omega_T}\left[1-\eta^2(n_g+n_e+1)\right]\nonumber\\
&+&\frac{\eta^2\Omega_R^2}{4\omega_T}\sum_{\substack{k=\pm 1\\n_g-n_e+k\neq 0}} \frac{n_g+n_e+1}{n_g-n_e+k}
+\mathcal{O}(\eta^4),
\end{eqnarray}
which is valid for $n_g\neq n_e$, since it follows directly from (\ref{VBS}) that the position of the central 
(carrier, $n_g=n_e$) resonances  are not shifted to any order in $\eta$: 
the shifts from both sides (positive and negative) detunings are compensated and canceled. (Contrast this with \cite{LME07}, where a different type of shift of the carrier was studied, defined by the excitation peak position rather than by the 
level structure). 

The first line in Eq. (\ref{VBS_LD}) is nothing but the (off-resonant) carrier contribution to the
vibrational Bloch-Siegert shift, while the second line is the contribution of the adjacent sideband transitions. It is remarkable
that in the LD-regime the main contribution to any sideband frequency shift is always given by the coupling to off-resonant carrier transitions, 
no matter how far the sideband is from the carrier. 

Note that in the strict LD ($\eta=0$) limit, the states $|g,n_g\ra$ and $|e,n_e\ra$ are not connected by the perturbation 
$V$, and $R_{ge}$ will be zero to all orders. 
The energy levels will cross while being shifted due to the effect of the carrier transitions, which are the only 
surviving contributions if $\eta=0$. The two levels are Stark shifted in opposite directions by $R_{gg}=-R_{ee}=\hbar\Omega_R^2/4\Delta_0$, shifting the position of the resonance by $-\frac{\Omega_R^2}{2\Delta_0}$
see Fig. \ref{zoomed_level_fig} (thick dashed line).


The vibrational Bloch-Siegert shifts can be seen experimentally and are indeed quite relevant.   
In experiments with single trapped ions to implement quantum gates \cite{steane00,kaler03,SchmidtKaler03},
ground state sideband cooling \cite{EMSB03} or quantum state engineering \cite{LBMW03,MMKIW96,roos99}, the laser has to be  
tuned precisely to a given sideband, so the shift has to be taken into account. 
For recent experiments with $^{40}$Ca$^+$ ions \cite{Riebe06} ($\omega_T=2\pi\times1.36$ MHz,
 $\Omega_R=2\pi\times53$ kHz and $\eta=0.083$),
the displacement of the resonant frequency at the first blue sideband is as high as one kHz, which 
corresponds to a relative displacement of about $10^{-3}$, 
see Fig. \ref{barplot}.

%
\section{Discussion}
\subsection{Comparison with previous works}
We have shown that the effect of the vibrational counter-rotating terms is to shift the position of the sidebands, an effect that may be related to the ordinary Bloch-Siegert shift. 
Compared to previous results in \cite{steane00,aniello04}, our formulation is 
more general and applies to arbitrary sidebands.  
In \cite{steane00}, Steane {\it et al} describe the  shift on the first red sideband as the result of the light shifts of the levels because of off-resonant transitions.
Aniello {\it et al} \cite{aniello04}, after a series of unitary transformations
of the Hamiltonian end up with a resonance condition which gives a shift also for the first red sideband. In both
cases the calculated shift is given by
\begin{equation}
\label{falta}
\delta\omega=\frac{\Omega_R^2}{2\omega_T}+\frac{\eta^2\Omega_R^2}{4\omega_T},
\end{equation}
$(n_g=1, n_e=0)$, which differs from our result in (\ref{VBS_LD}) 
by the $\eta^2$ correction from off-resonant carrier transitions.
This correction is indeed easy to miss (in particular, it is  
overlooked if $\la n|e^{i\eta(a+a^\dagger)}|n\ra$ is approximated 
by keeping only terms up to linear order in $\eta$), but it becomes quite significant  
at moderate values of $\eta$, as shown by        
the numerical comparison of the exact result (from diagonalization of the Hamiltonian with a large basis), and the shifts given by Eqs. (\ref{VBS_LD})
and (\ref{falta}) in Fig. \ref{vbs_eta_fig}(b). 
\subsection{RWA vs. vibrational RWA}
\label{RWAvsVRWA}

In order to highlight the differences between the \emph{usual} RWA and the VRWA (and thus the differences
between the \emph{usual} Bloch-Siegert shift and its vibrational  version), we shall now 
compare our trapped ion system with 
the interaction of a two-level atom and a quantized 
field mode of frequency $\omega$. These two
systems are in many ways similar if the ion is assumed to be trapped within the LD regime \cite{BWR92,CBP93}.
Since the RWA has to do with co-rotating and counter-rotating terms
in a time dependent Hamiltonian, the different nature of both RWA's is more clearly understood
if the Hamiltonians describing both systems are written in an interaction picture where the free evolution
of the amplitudes is removed, i. e., 
\begin{eqnarray}
\label{field_ion_comp}
H_F(t)&=&\frac{\hbar\Omega_R}{2}\left[\left(ae^{-i\omega t}+a^\dag e^{i\omega t}\right)e^{i\omega_0t}\sigma_++H.c.\right],
\nonumber\\
H_{TA,LD}(t)&=&\frac{\hbar\Omega_R}{2}\left[\left(1+i\eta ae^{-i\omega_T t}+i\eta a^\dag e^{i\omega_T t}\right)
e^{-i\Delta t}\sigma_+\right.
\nonumber\\
&+&H.c.\big],
\end{eqnarray}
where the subscript $F$ refers to the atom-quantized field case, and $TA,LD$ to the trapping case, see  Eq.(\ref{time_dep_ham_02}), in the Lamb-Dicke regime.\footnote{
The simplified Hamiltonian $H_{TA,LD}(t)$ would not give 
the correct $\eta^2$ correction to the vibrational Bloch-Siegert 
shift from off-resonant carrier transitions,
see Eq. (\ref{VBS_LD}), but for the qualitative 
comparison of both systems this approximation is enough.}
The creation and annihilation operators have different meaning depending on the system: they 
increase or decrease the number of photons of a given Fock state in the first case or they 
add or remove a vibrational quantum in the trapped ion case. 
Despite these conceptual differences, both systems look formally very similar. 

If near resonant processes are considered ($\omega\sim\omega_0$) in $H_F$, counter-rotating terms like $\sigma_-a$ and $\sigma_+a^\dag$ can be 
neglected since they oscillate with a frequency $\sim2\omega$, much faster than the co-rotating terms, whose
frequency of oscillation is $\delta=\omega-\omega_0$.
Counter-rotating terms can never be resonant since both the mode frequency $\omega$ and the transition frequency 
$\omega_0$ are positive quantities.
The main effect of keeping these counter-rotating terms is a shift on the apparent 
resonant frequency of the atom $\omega_0$, i. e., the ordinary Bloch-Siegert shift \cite{cohen73}.

In the trapped ion case, the role of $\omega_0$ is played by the 
frequency difference or detuning $\Delta=\omega_L-\omega_0$,
which may happen to be positive (blued detuned) or negative (red detuned).
Terms like $\sigma_+a$ will be resonant if the laser is tuned to the first red sideband. This 
is similar to the atom-field coupling case, since after all the interaction is described by a 
Jaynes-Cummings type Hamiltonian. 
However terms like $\sigma_+a^\dag$, which would be counter rotating
in the atom-field coupling case, are resonant in the trapped ion system when the laser is tuned to the first blue 
sideband and the system is approximately described by an anti
Jaynes-Cummings Hamiltonian.

Another important difference between these systems is that for  
the atom in the quantized-field, transitions where only the internal state of the atom changes (with no photon 
absorption or emission) do not happen, while in the trapped ion case these carrier-type transitions
give in fact the main contribution of the LD expansion, i.e.,
the zeroth order in $\eta$ contribution in 
Eq. (\ref{time_dep_ham_01}). 
When the laser is tuned to a given blue or red sideband, these 
carrier terms become counter-rotating and are responsible for the leading order contribution in $\eta$
of the vibrational Bloch-Siegert shift, Eq. (\ref{VBS_LD}), associated with 
the Stark shift of the energy levels as we have previously pointed out.
This also explains a factor of two discrepancy in the expressions of the dominant 
terms: $\delta\omega_{BS}=\Omega_R^2/4\omega$ for the standard Bloch-Siegert shift \cite{AllenEberly}, whereas $|\delta\omega|=\Omega_R^2/2\omega_T$ for the vibrational Bloch-Siegert shift of the first red or blue sidebands.  
At variance with the vibrational effect the ordinary Bloch-Siegert shift is too small to have been observed for optical transitions because of the large frequency in the denominator, whereas it is relatively easy to observe in the radio-frequency 
domain \cite{HFSSL07}.  

\begin{acknowledgments}
We thank E. Solano for commenting on the manuscript. 
This work has been supported by Ministerio de Educaci\'on y Ciencia 
(FIS2006-10268-C03-01, CSD2006-00019, FIS2005-08257, FIS2007-66944, CSD2006-00019 Consolider Project), by 
the UPV-EHU (00039.310-15968/2004), and the European Commission 
``EMALI'', MRTN-CT-2006-035369; ``SCALA'', Contract No. 015714.
\end{acknowledgments}

\appendix
%
\section{Calculation of matrix elements of the level shift operator}
\label{matrix_elements_calculation}
In this Appendix the matrix elements of the levels shift operator are explicitly calculated to leading order in the
perturbation $V$, see Eq. (\ref{perturbation}). 
We will consider the crossing between the 
$|g,n_g\ra$ and $|e,n_e\ra$ bare states, with $P=|g,n_g\ra\la g,n_g|+|e,n_e\ra\la e,n_e|$ and $Q=1-P$.
The first order term $\la g,n_g|V|g,n_g\ra$ vanishes, since the perturbation
does not connect states with the same internal atomic state. The next order will be given by
\begin{eqnarray}
R_{gg}(E_0)&=&\la g,n_g| V\frac{Q}{E_0-H_B}V|g,n_g\ra
\nonumber\\
&=&\frac{\hbar\Omega_R}{2}\la g,n_g|V\frac{Q}{E_0-H_B}e^{i\eta(a+a^\dag)}|e,n_g\ra
\nonumber\\
&=&\frac{\hbar\Omega_R}{2}\sum_{k\ne n_e} \frac{\chi_{n_g,k}}{E_0-E_{e,k}^{(B)}} \la g,n_g|V|e,k\ra
\nonumber\\
&=&\left(\frac{\hbar\Omega_R}{2}\right)^2\sum_{k\ne n_e} \frac{\left|\chi_{n_g,k}\right|^2}{E_0-E_{e,k}^{(B)}}
\nonumber\\
&=&\sum_{k\neq n_e}\frac{\left|\hbar\Omega_{n_g,k}/2\right|^2}{E_0-E_{e,k}^{(B)}},
\end{eqnarray}
where  
\begin{eqnarray}
\chi_{nn'}&=&\la n|e^{i\eta(a+a^\dag)}|n'\ra
\nonumber\\
&=&
e^{-\eta^2/2}\left(i\eta\right)^{|n-n'|}\!\sqrt{\frac{n_{_<}!}{n_{_>}!}}
L_{n_{_<}}^{|n-n'|}(\eta^2),
\end{eqnarray}
$n_{_{<}}$ ($n_{_>}$) being the lesser (greater) of $n$ and $n'$ and 
$L_n^{\alpha}$ the generalized Laguerre functions, 
\begin{equation}
L_n^{\alpha}(X)=\sum_{k=0}^n (-1)^k 
\binom{n+\alpha}{n-k}
\frac{X^k}{k!}.
\end{equation}
In an analogous way, $R_{ee}$ is obtained:
\begin{equation}
R_{ee}(E_0)=\sum_{k\neq n_g}\frac{\left|\hbar\Omega_{n_e,k}/2\right|^2}{E_0-E_{g,k}^{(B)}}.
\end{equation}


\begin{thebibliography}{1}

\bibitem{BS40} F. Bloch and A. Siegert, 
Phys. Rev.  {\bf 57}, 522 (1940)

\bibitem{AllenEberly} L. Allen and J. H. Eberly
{\it Optical Resonance and two-level atoms}, 
Dover Publications, Inc., New York (1975)

\bibitem{WWM97} C. Wei {\it et al.}
J. Phys. B  {\bf 30}, 4877-4888 (1997)

\bibitem{LBMW03} D. Leibfried, R. Blatt, C. Monroe, and D. Wineland, 
Rev. Mod. Phys. {\bf 75}, 281 (2003).

\bibitem{steane00}     A. Steane, C. F. Roos, D. Stevens, A. Mundt, D. Leibfried, F. Schmidt-Kaler, and R. Blatt
Phys. Rev. A {\bf 62}, 042305 (2000)  

\bibitem{SchmidtKaler03} F. Schmidt-Kaler {\it et al.},
Appl. Phys. B  {\bf 77}, 789-796 (2003)

\bibitem{Orszag} M. Orszag, {\it Quantum Optics} (Springer, Berlin, 2000).

\bibitem{LM07_b} I. Lizuain and J. G. Muga, 
Phys. Rev. A{\bf 75}, 033613 (2007)

\bibitem{CDG98} C. Cohen-Tannoudji, J. Dupont-Roc and G. Grynberg,
{\it Atom-Photon Interactions} Wiley Science Paperback Series (1998)

\bibitem{cohen73} C. Cohen-Tannoudji, J. Dupont-Roc, and C. Fabre,
Phys. B: At. Mol. Phys. {\bf 6}, L214-L217 (1973)

\bibitem{WI79} D. J. Wineland and W. M. Itano,  Phys. Rev. A {\bf 20}, 1521 (1979)

\bibitem{wineland98} D. J. Wineland, C. Monroe, W. M. Itano, D. Leibfried, 
B. E. King, and D. M. Meekhof, J. Res. Natl. Inst. Stand. Technol. {\bf 103}, 259 (1998).

\bibitem{aniello04} P. Aniello {\it et al.}, 
Journal of Russian Laser Research, Volume 25, Number 1 (2004).

\bibitem{LME07} I. Lizuain, J. G. Muga, and J. Eschner,
Phys. Rev. A {\bf 76}, 033808 (2007). 

\bibitem{kaler03} F. Schmidt-Kaler {\it et al.},
Nature (London) {\bf 422}, 408 (2003). 

\bibitem{EMSB03} J. Eschner {\it et al.}, 
J. Opt. Soc. Am. B {\bf 20}, 1003 (2003).

\bibitem{MMKIW96} D. M. Meekhof, C. Monroe, B. E. King, W. M. Itano,
and D. J. Wineland,
Phys. Rev. Lett. {\bf 76}, 1796 (1996).

\bibitem{roos99} Ch. Roos, Th. Zeiger, H. Rohde, H. C. N\"agerl, J. Eschner, D. Leibfried, F. Schmidt-Kaler, and R. Blatt
Phys. Rev. Lett. {\bf 83}, 4713 (1999).

\bibitem{Riebe06}  M. Riebe, K. Kim, P. Schindler, T. Monz, P. O. Schmidt, T. K. K\"orber, W. H\"ansel, H. H\"affner, 
C. F. Roos, and R. Blatt,
Phys. Rev. Lett. {\bf 97}, 220407 (2006).


\bibitem{BWR92} C. A. Blockley, D. F. Walls and H. Risken, 
Europhys. Lett. {\bf 17}, 509 (1992)

\bibitem{CBP93}  J. I. Cirac, R. Blatt, A. S. Parkins, and P. Zoller,
Phys. Rev. Lett. {\bf 70}, 762 (1993)


\bibitem{HFSSL07} S. Hofferberth, B. Fischer, T. Schumm, J. Schmiedmayer, and I. Lesanovsky,
Phys. Rev. A {\bf 76}, 013401 (2007).







\end{thebibliography}
\end{document}